\documentclass[5pt]{article}
\usepackage{amsmath}
\usepackage[latin1]{inputenc}
\usepackage{amsmath,amsthm,amssymb}
\usepackage{graphicx}
\usepackage{color}
\usepackage{multirow}
\usepackage{rotating}
\usepackage{subeqnarray}
\usepackage{cases}
\usepackage{cancel}

\renewcommand{\theequation}{\thesection.\arabic{equation}}

\title{ Further Results on Asymmetric Single Correcting Codes of  Magnitude Four}
\author{{Derong Xie  \qquad\quad    Jinquan Luo}\footnote{ The authors are with school of mathematics and
statistics \& Hubei Key Laboratory of Mathematical Sciences,
Central China Normal University
Wuhan 430079, China.
E-mail: luojinquan@mail.ccnu.edu.cn(J.Luo); derongxie@yahoo.com.
}}
\date{}
\begin{document}
\baselineskip15pt \maketitle
\renewcommand{\theequation}{\arabic{section}.\arabic{equation}}
\catcode`@=11 \@addtoreset{equation}{section} \catcode`@=12

$\mathbf{Abstract-}$An error model with asymmetric single error with magnitude four is considered. In this paper, the constructions of codes correcting single error of magnitude four over $\mathbb{Z}_{2^{a}3^{b}r}$ are studied which is equivalent to construct $B_{1}[4](2^{a}3^{b}r)$ sets. Firstly, we reduce the construction of a maximal size $B_{1}[4](2^{a}3^{b}r)$ set for $a\geq4$ and $\gcd(r,6)=1$ to the construction of a maximal size $B_{1}[4](2^{a-3}3^{b}r)$ set. Furthermore, we will show that maximal size $B_{1}[4](8\cdot3^{b}r)$ sets can be reduced to maximal size $B_{1}[4](3^{b}r)$ sets. Finally, we give lower bounds of maximal size $B_{1}[4](2r)$ and $B_{1}[4](2\cdot3^{b}r)$ sets.

\emph{$\mathbf{Index \ Terms-}$}Asymmetric error,\  single error,\ flash memories,\ limited magnitude error.

\section{Introduction}

The asymmetric channel with limited magnitude errors was introduced in \cite{AAK} and it can be applied to some multilevel flash memories. Nonsystematic and systematic codes correcting asymmetric errors of given maximum magnitude are presented in \cite{AAKT1,EB}. In the asymmetric error model, a symbol $a$ over an alphabet
$$\mathbb{Z}_{q}=\{0,1,\cdots,q-1\}$$
may be modified during transmission into $b$, where $b\geq a$, and the probability that $a$ is changed to $b$ is considered to be the same for all $b>a$. For some applications, the error magnitude $b-a$ is not likely to exceed a certain level $\lambda.$  In general, the errors are mostly asymmetric and some classes of construction of codes correcting such errors were studied in \cite{CSB,KBN,KLNY}. Also, several constructions of systematic codes correcting single errors are given in \cite{KBN}
and the symmetric case is closely related to equi-difference conflict-avoiding codes, see e.g., \cite{LMJ,LMS,MSJ,KS, WF}. In addition, splitter sets can be regarded as codes correcting single limited magnitude errors in flash memories, see e.g., \cite{BE,M,YKB,ZG,ZZG}.

We briefly recall known links between codes correcting single errors and $B_{1}[\lambda](q)$ sets which have been introduced in \cite{KBN,KLNY,KLY}. For integers $k_{1},k_{2}$, where $k_{1}\leq k_{2}$, we use the notation
$$[k_{1},k_{2}]=\{k_{1},k_{1}+1,k_{1}+2,\cdots,k_{2}\}.$$
If $H$ is an $h\times m$ matrix over $\mathbb{Z}_{q}$, the corresponding code of length $m$ with parity check matrix $H$, is
$$\mathcal{C}_{H}=\{\mathbf{x}\in\mathbb{Z}_{q}^{m} \ | \ \mathbf{x}H^{t}=\mathbf{0}\}$$
where $H^{t}$ denotes the transpose of $H$.

Let $\mathcal{E}\subset\mathbb{Z}_{q}^{m}$ be the set of error patterns that we want to
correct and consider single errors of magnitude at most $\lambda$. If $\mathbf{x}\in \mathcal{C}_{H}$ is a sent codeword and $\mathbf{e}\in\mathcal{E}$ is an error introduced during transmission, then the received $m$-tuple is $\mathbf{y}=\mathbf{x}+\mathbf{e}$. Therefore
$$\mathbf{y}H^{t}=\mathbf{x}H^{t}+\mathbf{e}H^{t}=\mathbf{e}H^{t}.$$
As usual, $\mathbf{e}H^{t}$ is called the syndrome of $\mathbf{e}$. Let
$$\mathcal{S}_{H,\mathcal{E}}=\{\mathbf{e}H^{t} \ | \mathbf{e}\in\mathcal{E}\}$$
be the set of syndromes. We require all of $\mathbf{e}H^{t}$ to be distinct for $\mathbf{e}\in\mathcal{E}$, i.e., $|\mathcal{S}_{H,\mathcal{E}}|=|\mathcal{E}|$. In this case, the code is able to correct all error patterns in $\mathcal{E}$. In particular,
$$\bigcup_{\mathbf{x}\in\mathcal{C}_{H}}\{\mathbf{x}+\mathbf{e} \ \big| \ \mathbf{e}\in\mathcal{E}\}$$
is a disjoint union.

For $h=1$, that is, $H=(b_{0},b_{1},\cdots,b_{m-1})$. The error patterns we consider are
$\mathcal{E}_{\lambda,m}$, the set of sequences $(e_{0},e_{1},\cdots,e_{m-1})\in [0,\lambda]^{m}$ of Hamming weight at most 1.
Consider
$$B=\{b_{0},b_{1},\cdots,b_{m-1}\}$$
of distinct positive integers such that the corresponding syndromes
$$\mathcal{S}=\left\{\sum_{j=0}^{m-1}e_{j}b_{j} \ (\mbox{mod}\ q)\ \Big| \ (e_{0},e_{1},\cdots,e_{m-1})\in\mathcal{E}_{\lambda,m}\right\}$$
are distinct. The set $B$ is called a $B_{1}[\lambda](q)$ set, see \cite{KBN}. In other words, for any two different integers $i,j\in [1, \lambda]$,
\begin{equation}\label{Bset}
|B|=|i\cdot B| \qquad \text{and} \qquad i\cdot B\cap j\cdot B=\emptyset
\end{equation}
where $i\cdot B=\{ib \ | \ b\in B\}.$
The corresponding code is
$$\mathcal{C}_{B}=\left\{(x_{0},x_{1},\cdots,x_{m-1})\in\mathbb{Z}_{q}^{m} \ \left|\ \sum_{i=0}^{m-1}x_{i}b_{i}\equiv0 \ (\mbox{mod}\ q)\right.\right\}.$$

 The construction of a maximal size $B_{1}[3](2^{a}r)$ set, $B_{1}[3](3^{b}r)$ set and $B_{1}[4](3^{b}r)$ set can be found in \cite{KLNY}. Several construction of maximal size $B_{1}[4](2^{a}r)$ sets can be found in \cite{XL}. In this paper, we will discuss maximal size $B_{1}[4](2^{a}3^{b}r)$ sets with $\gcd(r,6)=1$. In Section \uppercase\expandafter{\romannumeral2}, we reduce the construction of a maximal size $B_{1}[4](2^{a}3^{b}r)$ set for $a\geq4$ to the construction of a maximal size $B_{1}[4](2^{a-3}3^{b}r)$ set. In Section \uppercase\expandafter{\romannumeral3}, we consider the construction of a maximal size $B_{1}[4](8\cdot3^{b}r)$ set. In section \uppercase\expandafter{\romannumeral4}, a lower bound of maximal size $B_{1}[4](12r)$ set is given from a maximal size $B_{1}[4](2r)$ set. In Section \uppercase\expandafter{\romannumeral5}, we give a lower bound of maximal size $B_{1}[4](2\cdot3^{b}r)$ sets. Finally, we give a short summary of this paper in Section \uppercase\expandafter{\romannumeral6}.

\section{Maximal size $B_{1}[4](2^{a}3^{b}r)$ set}

For integer $q=2^{a}3^{b}r$ with $\gcd(r,6)=1$, we will reduce the construction of a maximal size $B_{1}[4](2^{a}3^{b}r)$ set for $a\geq4$ to the construction of a maximal size $B_{1}[4](2^{a-3}3^{b}r)$ set. We firstly describe the following notations.

Let $\mathbb{Z}_{d}^{*}=\{x\in \mathbb{Z}_{d} \ | \ \gcd(x,d)=1\}$. For any positive integer $l$ coprime to $d$, let $\mbox{ord}_{d}(l)$ be the order of $l$ in $\mathbb{Z}_{d}^{*}$, that is,
$$\mbox{ord}_{d}(l)=\min\{n>0 \ | \ l^{n}\equiv1 \ (\mbox{mod}\ d)\}.$$

For $m\in\mathbb{Z}_{d}^{*}$ and $d$ a divisor of $q$, we let $\beta=mq/d.$ For $\gcd(ml,d)=1$, define the cyclotomic set
$$\sigma_{l}(\beta)=\{l^{i}\beta \ (\mbox{mod}\ q) \ |\ i\geq 0 \}.$$
Then $|\sigma_{l}(\beta)|=\mbox{ord}_{d}(l).$

For integer $a$, $(a \ \text{mod} \ q)$ denotes the least non-negative residue of $a$ modulo $q$. Define $M_{4}(q)$ to be the maximal size of a $B_{1}[4](q)$ set.
For $0\leq i\leq a$, $0\leq j\leq b$, $\gcd(r,6)=1$ and $d\mid r$, let
\begin{align}
& V_{d}=\{xr/d \ (\mbox{mod} \ 2^{a}3^{b}r)\ | \ x  \in \mathbb{Z}_{2^{a}3^{b}d}, \ \gcd(x,d)=1\},\label{def Vd} \\
& U_{ij}=\{x\in \mathbb{Z}_{2^{a}3^{b}r} \ | \ \gcd(x,2^{a}3^{b})=2^{i}3^{j}\}\label{def Uij}.
\end{align}
Then
$$\mathbb{Z}_{2^{a}3^{b}r}=\bigcup_{d\mid r}V_{d}=\bigcup_{0\leq i\leq a}\bigcup_{0\leq j\leq b}U_{ij}.$$

Let
\begin{equation}\label{def M'}
M_{4}^{'}(2^{a}3^{b}d)=\max\limits_{B \ \mbox{is a} \ B_{1}[4](2^{a}3^{b}r) \ \mbox{set}}\left|B\cap V_{d} \right|.
\end{equation}
Then
$$M_{4}(2^{a}3^{b}r)=\sum_{d\mid r}M_{4}^{'}(2^{a}3^{b}d).$$

If $a\geq3$, we consider the following disjoint decomposition:
\begin{equation}\label{def L}
\mathbb{Z}_{2^{a}3^{b}r}\backslash\{0\}=L_{0}\cup L_{1}\cup L_{2}\cup L_{3},
\end{equation}
where
$$L_{i}=\{2^{i}x \ \mbox{mod}\ 2^{a}3^{b}r \mid 1\leq x\leq 2^{a-i}3^{b}r, 2\nmid x\} \ \mbox{for} \ 0\leq i \leq2,$$
and
$$L_{3}=\{8x \ \mbox{mod}\ 2^{a}3^{b}r \mid 1\leq x< 2^{a-3}3^{b}r\}.$$

$Theorem \ 1:$ If $a\geq 4$ and $\gcd(r,6)=1$, then
$$M_{4}(2^{a}3^{b}r)=M_{4}(2^{a-3}3^{b}r)+2^{a-3}3^{b}r.$$
$Proof:$ Note that $a\geq 4$. Let
\begin{align*}
& S_{1}=\left\{4i+1 \ | \ 0\leq i<2^{a-4}3^{b}r \right\},\\
& S_{2}=\left\{4i+3+2^{a-2}3^{b+1}r \ | \ 0\leq i<2^{a-4}3^{b}r \right\}.
\end{align*}
Let $S_{3}$ be a $B_{1}[4](2^{a-3}3^{b}r)$ set. Denote by $S=S_{1}\cup S_{2}$ and $B=S\cup S^{'} \ \mbox{with} \ S^{'}=\{8c \ (\mbox{mod}\ 2^{a}3^{b}r) \mid c\in S_{3}\}$. In the following, we will verify $B$ satisfies (1.1) and so $B$ is a $B_{1}[4](2^{a}3^{b}r)$ set. Since $S^{'}$ is a $B_{1}[4](2^{a}3^{b}r)$ set contained in $L_{3}$ which implies that we just need to prove $S$ is a $B_{1}[4](2^{a}3^{b}r)$ set.

Clearly, $|S_{1}|=|S_{2}|=2^{a-4}3^{b}r$ and $S_{1}\cap S_{2}=\emptyset$ which implies $|S|=2^{a-3}3^{b}r$. For $0\leq i<2^{a-4}3^{b}r$, we obtain
\begin{align*}
& 0\leq8i+2<2^{a}3^{b}r \ \mbox{and} \ 8i+2\equiv2 \ (\mbox{mod} \ 8),\\
& 8i+6+2^{a-1}3^{b+1}r (\mbox{mod} \ 2^{a}3^{b}r)=8i+6+2^{a-1}3^{b}r \ \mbox{and} \ 8i+6+2^{a-1}3^{b}r\equiv6 \ (\mbox{mod} \ 8).
\end{align*}
Then $|2S|=2^{a-3}3^{b}r$. Similarly, $|3S|=|4S|=2^{a-3}3^{b}r$.

Note that $S\subset L_{0}$, $2S\subset L_{1}$, $3S\subset L_{0}$ and $4S\subset L_{2}$. Therefore, in order to show that $iS\cap jS=\emptyset$ for distinct $i,j\in[1,4]$, it suffices to prove $S\cap3S=\emptyset$. For any
$$x\in3S=\left\{12i+3,\ 12i+9+2^{a-2}3^{b}r \ | \ 0\leq i<2^{a-4}3^{b}r \right\},$$
either
$$x\equiv1 \ (\mbox{mod}\ 4) \qquad \text{and} \qquad 2^{a-2}3^{b}r<x<q$$
or
$$\quad \; x\equiv3 \ (\mbox{mod}\ 4) \qquad \text{and} \qquad 0<x<2^{a-2}3^{b+1}r.$$
Hence $S\cap3S=\emptyset$.
Then $B$ is a $B_{1}[4](2^{a}3^{b}r)$ set of size $M_{4}(2^{a-3}3^{b}r)+2^{a-3}3^{b}r$.

Finally we will show $B$ is maximal. Note that at least one of $x,2x,3x,4x$ belongs to $L_{2}$ for any $x\in L_{0}\cup L_{1}\cup L_{2}$. Therefore, in $L_{0}\cup L_{1}\cup L_{2}$, at most $|L_{2}|=2^{a-3}3^{b}r$ elements can be chosen in a $B_{1}[4](2^{a}3^{b}r)$ set.
Also, none of $x,2x,3x,4x$ belong to $L_{0}\cup L_{1}\cup L_{2}$ for any $x\in L_{3}$. Since
$$(S\cup 2S\cup 3S\cup 4S)\subset (L_{0}\cup L_{1}\cup L_{2}),$$
then the set $B$ is a maximal size $B_{1}[4](2^{a}3^{b}r)$ set of size $M_{4}(2^{a-3}3^{b}r)+2^{a-3}3^{b}r$.\hfill$\blacksquare$

$Example \ 1:$
\begin{itemize}
\item For $q=48$, we have
$$M_{4}(48)= M_{4}(6)+6.$$
It is easy to check that $\{1\}$ is a maximal size $B_{1}[4](6)$ set. The construction of a maximal size $B_{1}[4](48)$ set in the proof of Theorem 1 is presented as follows. Firstly, $S_{1}=\{1,5,9\}$ and $S_{2}=\{39,43,47\}.$

Therefore
$$B=S_{1}\cup S_{2}\cup \{8\cdot1\}=\{1,5,8,9,39,43,47\}$$
is a maximal size $B_{1}[4](48)$ set with size 7.
The corresponding code is
   $$\mathcal{C}_{B}=\left\{(x_{0},x_{1},\cdots,x_{6})\in\mathbb{Z}_{48}^{7} \ \big|\ x_{0}+5x_{1}+8x_{2}+9x_{3}+39x_{4}+43x_{5}+47x_{6}\equiv0 \ (\mbox{mod}\ 48)\right\}.$$
   Set the codeword $\mathbf{c}=(5,3,2,3,1,1,1)$.  An error $\mathbf{e}=(0,0,0,0,2,0,0)$ occurs during transmission and so the received vector $\mathbf{y}=\mathbf{c}+\mathbf{e}=(5,3,2,3,3,1,1)$. We calculate
   $$1\cdot5+5\cdot3+8\cdot2+9\cdot3+39\cdot3+43\cdot1+47\cdot1\equiv30 \ (\mbox{mod}\ 48).$$
   The syndrome 30 can be uniquely represented as $b_{i}\cdot m\equiv30(\mbox{mod}\ 48)$ for some $b_{i}\in B$ and $m\in[1,4]$ (actually $b_{i}=39$, $m=2$ and $i=5$). Furthermore, the indice $i$ indicates the error location. Hence error vector $$\mathbf{e}=(0,0,0,0,2,0,0).$$
   Therefore,
   $$\mathbf{c}=\mathbf{y}-\mathbf{e}=(5,3,2,3,1,1,1).$$
\item For $q=240$, we have
$$M_{4}(240)= M_{4}(15)+30.$$
By  Theorem 3, the set $\{1,7\}$ is a maximal size $B_{1}[4](15)$ set.
The construction of $B_{1}[4](240)$ set in the proof of Theorem 1 is presented as follows. Firstly,
\begin{align*}
& S_{1}=\{1, 5, 9, 13, 17, 21, 25, 29, 33, 37, 41, 45, 49, 53, 57\}.\\
& S_{2}=\{183, 187, 191, 195, 199, 203, 207, 211, 215, 219, 223, 227, 231, 235, \
239\}.
\end{align*}
Therefore
\begin{align*}
& B=S_{1}\cup S_{2}\cup \{8\cdot1,8\cdot7\}=\{1, 5, 8, 9, 13, 17, 21, 25, 29, 33, 37, 41, 45, 49, 53, 56, 57, 183, 187, 191,\\
& 195, 199, 203, 207, 211, 215, 219, 223, 227, 231, 235, 239\}
\end{align*}
is a maximal size $B_{1}[4](240)$ set with size 32.
\end{itemize}

\section{Maximal size $B_{1}[4](8\cdot3^br)$ set}

Unfortunately, the technique to construct $B_{1}[4](2^{a}3^{b}r)$ set in Section \uppercase\expandafter{\romannumeral2} is not applicable for $a\leq3$. In this section, we reduce the construction of a maximal size $B_{1}[4](8\cdot3^{b}r)$ set to the construction of a maximal size $B_{1}[4](3^{b}r)$ set. For $b\geq2$, the maximal size $B_{1}[4](3^{b}r)$ set is given in [\cite{KLNY}, Th.10]. We give an explicit construction of a maximal size $B_{1}[4](3r)$ set in Theorem 3.

$Theorem \ 2:$ If $b\geq1$ and $\gcd(r,6)=1$, then
$$M_{4}(8\cdot3^{b}r)=M_{4}(3^{b}r)+3^{b}r.$$
$Proof:$ For $0\leq j \leq b$ and $d\mid r$, let $\alpha=r/d$ and
\begin{eqnarray*}
S_{j,d}\!&=&\!\left\{i\cdot3^{b-j}\alpha+2\cdot3^{b}r \ | \ 1\leq i\leq1+3^{j}d, \ \gcd(i,2\cdot3^{j}d)=1 \right\}\\
&&\cup\left\{i\cdot3^{b-j}\alpha \ | \ 1+3^{j}d< i<2\cdot3^{j}d, \ \gcd(i,2\cdot3^{j}d)=1 \right\}.
\end{eqnarray*}
Note that
$$|S_{j,d}|=
\left\{
\begin{array}{ll}
\varphi(d)  ,&  j=0;\\
2\cdot3^{j-1}\varphi(d) ,&  1\leq j \leq b.
\end{array}
\right.
$$
Denote by $S=\bigcup_{d|r}\bigcup_{j=0}^{b}S_{j,d}$. For any $x\in S_{j,d}$,
\begin{itemize}
\item if $j=0$, then $3^{b}\mid x;$
\item if $1\leq j \leq b$, then $3^{b-j}\mid x \ \mbox{and} \ 3^{b-j+1}\nmid x.$
\end{itemize}
As a consequence, $S_{j,d}\cap S_{j^{'},d^{'}}=\emptyset$ for $j\neq j^{'}$.
It is easy to see that $S_{j,d}\cap S_{j^{'},d^{'}}=\emptyset$ for $d\neq d^{'}$.
Therefore, if $(d,j)\neq(d^{'},j^{'})$, then
$$S_{j,d}\cap S_{j^{'},d^{'}}=\emptyset$$
and so
$$|S|=\sum_{d|r}\sum_{j=0}^{b}|S_{j,d}|=\sum\limits_{d|r}3^{b}\varphi(d)=3^{b}r.$$

Let $S_{b}$ be a $B_{1}[4](3^{b}r)$ set. Define
$$B=S\cup S^{'} \ \mbox{with} \ S^{'}=\{8c \ (\mbox{mod}\ 8\cdot3^{b}r) \mid c\in S_{b}\}.$$
In the following, we will verify $B$ satisfies (1.1) and so $B$ is a $B_{1}[4](2^{a}3^{b}r)$ set.
It is easy to verify that there do not exist distinct elements $x,y\in B$ such that $2x\equiv2y \ (\mbox{mod}\ 2^{k}r)$, $3x\equiv3y \ (\mbox{mod}\ 2^{k}r)$ or $4x\equiv4y \ (\mbox{mod}\ 2^{k}r)$. Hence all of $2B$, $3B$ and $4B$ have the same size as $B$.

Similarly to the analysis of Theorem 1, we just need to prove $S\cap3S=\emptyset$. Firstly, if $d\neq d^{'}$ or $j^{'}\neq j-1$, then
$$3S_{j,d}\cap S_{j^{'},d^{'}}=\emptyset.$$

Also, for any $y\in S_{(j-1),d}$ and $x\in S_{j,d}$, one has
$$3^{b-j+1}\alpha+3^{b}r<y\leq 3^{b-j+1}\alpha+3^{b+1}r,$$
and
$$(3x \ \text{mod} \ 8\cdot3^br)\in[1,3^{b-j+1}\alpha+3^{b}r]\cup(3^{b-j+1}\alpha+3^{b+1}r,8\cdot3^{b}r].$$
Hence $S_{(j-1),d}\cap3S_{j,d}=\emptyset$.

Finally,
\begin{eqnarray*}
S_{0,d}\!&=&\!\left\{i\cdot3^{b}\alpha+2\cdot3^{b}r \ | \ 1\leq i\leq d-1, \ \gcd(i,2d)=1 \right\}\\
&&\cup\left\{i\cdot3^{b}\alpha \ | \ 2+d\leq i\leq2d-1, \ \gcd(i,2d)=1 \right\}.
\end{eqnarray*}
For any $z\in S_{0,d}$, we can deduce
\begin{itemize}
\item $z\in [2\cdot3^{b}\alpha+3^{b}r,3^{b+1}r-3^{b}\alpha]$,
\item $(3z \ \text{mod} \ 8\cdot3^br)\in[1,3^{b}r-3^{b+1}\alpha]\cup[2\cdot3^{b+1}\alpha+3^{b+1}r,8\cdot3^{b}r]$
\end{itemize}
and so $S_{0,d}\cap3S_{0,d}=\emptyset$. Hence $S\cap3S=\emptyset.$
Then $B$ is a $B_{1}[4](8\cdot3^{b}r)$ set of size $M_{4}(3^{b}r)+3^{b}r$.

Similarly to the proof of Theorem 1, $B$ is maximal.\hfill$\blacksquare$\\

$Lemma \ 1:$ (\cite{KLNY}, Lemma 5) \ a). For $d=p_{1}^{e_{1}}p_{2}^{e_{2}}\cdots p_{s}^{e_{s}}$ with $p_{i}$ distinct primes not dividing $l$,
$$\mbox{ord}_{d}(l)=\mbox{lcm}\left(\mbox{ord}_{p_{1}^{e_{1}}}(l),\mbox{ord}_{p_{2}^{e_{2}}}(l),\cdots,\mbox{ord}_{p_{s}^{e_{s}}}(l)\right).$$
b). If $p$ is a prime not dividing $l$ and $l^{\mbox{ord}_{p}(l)}-1=p^{\mu_{p}}a$, where $\gcd(a,p)=1$, then
$$
\mbox{ord}_{p^{k}}(l)=
\left\{
\begin{array}{ll}
\mbox{ord}_{p}(l) & if \ k\leq\mu_{p},\\
p^{k-\mu_{p}}\mbox{ord}_{p}(l) & if \ k>\mu_{p},  \ p=2 \ \mbox{and} \ l\equiv1(\mbox{mod}\ 4),\\
p^{k-\mu_{p}}\mbox{ord}_{p}(l) & if \ k>\mu_{p} \ \mbox{and} \ p>2,\\
2 & if  \ p=2,\ l\equiv3(\mbox{mod}\ 4) \ \mbox{and} \ k=2,3,\\
2^{k-2} & if \ p=2,\ l\equiv3(\mbox{mod}\ 4) \ \mbox{and} \ k>3.
\end{array}
\right.
$$
\\

$Lemma \ 2:$ For integer $d\geq5$ and $\gcd(d,6)=1$, there exists a set  $\Gamma_{3d}$ of coset representatives of the group generated by $2$ in $\mathbb{Z}_{3d}^{*}$ such that for any $i,i^{'}\in[0, \ \left\lfloor\mbox{ord}_{d}(2)/3\right\rfloor]$ and any two distinct $x,y\in\Gamma_{3d}$,
$$2^{3i}x\not\equiv2^{3i^{'}}y \ (\mbox{mod} \ d).$$

$Proof:$  By Lemma 1,
 $$\mbox{ord}_{3d}(2)=\mbox{lcm}(\mbox{ord}_{3}(2),\mbox{ord}_{d}(2))=\mbox{lcm}(2,\mbox{ord}_{d}(2)).$$
 Then it can be divided into two cases.

$\mathbf{Case \ 1}$: $\mbox{ord}_{d}(2)$ is odd. In this case, $\mbox{ord}_{3d}(2)=2\mbox{ord}_{d}(2)$ and the canonical homomorphism
$$\mathbb{Z}_{3d}^{*}/\langle2\rangle_{3d}\longrightarrow\mathbb{Z}_{d}^{*}/\langle2\rangle_{d}$$
is a group isomorphism and so $\Gamma_{3d}$ is also a set of coset representatives of the group generated by $2$ in $\mathbb{Z}_{d}^{*}$.

$\mathbf{Case \ 2}$: $\mbox{ord}_{d}(2)$ is even. In this case, $\mbox{ord}_{3d}(2)=\mbox{ord}_{d}(2)$. Let
$$\Lambda_{3d}=\{x_{1},x_{2},\cdots,x_{|\Lambda_{3d}|}\}$$
be a set of coset representatives of the group generated by $2$ in $\mathbb{Z}_{3d}^{*}$.
Firstly, $3\nmid x_{j}$ for $j\in[1, \ |\Lambda_{3d}|].$ Suppose that there exist $x_{1}, x_{2}, x_{3}\in\Lambda_{3d}$ such that
$$x_{1}\equiv x_{2}\equiv x_{3} \ (\mbox{mod} \ d).$$
Since $x_{i}\not\equiv 0 \ (\mbox{mod} \ 3)$ for $1\leq i\leq3$ , we may assume that $x_{1}\equiv x_{2} \ (\mbox{mod} \ 3)$, which yields
$$x_{1}\equiv x_{2} \ (\mbox{mod} \ 3d).$$
It is a contradiction.

Hence, there are at most two different elements $x_{1},x_{2}
\in\Lambda_{3d}$ such that
$$x_{1}\equiv x_{2} \ (\mbox{mod} \ d)$$
and so $x_{2}\equiv x_{1}+ld \ (\mbox{mod} \ 3d)$ with $l=1 \ \text{or} \ 2$  and $3\nmid x_{1}x_{2}.$

For distinct $x,y\in\Lambda_{3d}$ and for any $i,i^{'}\in[0, \ \left\lfloor\mbox{ord}_{d}(2)/3\right\rfloor]$,
\begin{itemize}
\item if $y\in (x+ld)\cdot\langle2\rangle_{3d}$ with $3\nmid (x+ld)$, then we replace $y$ with $2x+2ld$ and
$$2^{3i}x\not\equiv2^{3i^{'}}(2x+2ld) \ (\mbox{mod} \ d).$$
\item If $y\not\in (x+ld)\cdot\langle2\rangle_{3d}$ with $3\nmid (x+ld)$, then
$$2^{3i}x\not\equiv2^{3i^{'}}y \ (\mbox{mod} \ d).$$
Otherwise $y\equiv 2^{3i-3i'}x  \ (\mbox{mod} \ d) $, i.e.,
\begin{itemize}
  \item[(i)] $y\in x\cdot\langle2\rangle_{3d}$
  and it contradicts to $x$ and $y$ are distinct in $\Lambda_{3d}$.
  \item[(ii)] $y\in (x+ld)\cdot\langle2\rangle_{3d}$
  and it contradicts to $y\not\in (x+ld)\cdot\langle2\rangle_{3d}$.
\end{itemize}
\end{itemize}\hfill$\blacksquare$\\

Recall $M_{4}^{'}(3d)$ in $(\ref{def M'})$. Combining Lemmas 1 and 2, we get the following main result.\\

$Theorem \ 3:$ Let $n=\mbox{ord}_{d}(2)$. Then $M_{4}(3r)=\sum\limits_{d\mid r}M_{4}^{'}(3d)$ where
$$M_{4}{'}(3d)=
\left\{
\begin{array}{ll}
\left(\left\lceil\frac{2n}{3}\right\rceil-1\right)\cdot\frac{\varphi(d)}{n} & \mbox{if} \ n \ \mbox{is \ odd},\\
\left\lfloor\frac{n}{3}\right\rfloor\cdot\frac{2\varphi(d)}{n} & \mbox{if} \ n \ \mbox{is \ even}.
\end{array}
\right.$$

$Proof:$  For any $d\mid r$, let $n_{1}=\mbox{ord}_{3d}(2)$. By Lemma 1,
 $$n_{1}=\mbox{lcm}(\mbox{ord}_{3}(2),\mbox{ord}_{d}(2))=\mbox{lcm}(2,\mbox{ord}_{d}(2)).$$
 For any $x$, denote by $\eta=xr/d$ (hence the value of $\eta$ varies with $x$). Let $\Gamma_{3d}$ be a set of coset representatives of the group generated by $2$ in $\mathbb{Z}_{3d}^{*}$ satisfying Lemma 2.
 Clearly, $M_{4}^{'}(3)=0.$

 If $d\geq5$, then it can be divided into two cases.

$\mathbf{Case \ 1}$: $n$ is odd, then $n_{1}=2n$.
\begin{itemize}
\item[(i)] If $3\nmid n$, then we choose
$$T_{d}=\bigcup_{x\in \Gamma_{3d}}\left\{ 2^{3i}\eta \ (\mbox{mod}\ 3r)\ | \ 0\leq i <\left\lfloor\frac{2n}{3}\right\rfloor\right\}.$$
Suppose that
$$3\cdot2^{3i}\eta\equiv3\cdot2^{3i^{'}}\eta\ (\mbox{mod} \ 3r)$$
for distinct elements $2^{3i}\eta,\ 2^{3i^{'}}\eta \in T_{d}$ with $0\leq i,i^{'}<\lfloor2n/3\rfloor$. Then
$$2^{3i}\equiv2^{3i^{'}}\ (\mbox{mod} \ d)$$
which implies that $|i-i^{'}|=\frac{n}{3}$  and it contradicts to $3\nmid n.$ Therefore,
$$3\cdot2^{3i}\eta\not\equiv3\cdot2^{3i^{'}}\eta\ (\mbox{mod} \ 3r).$$
\item[(ii)] If $3\mid n$, then we choose
$$T_{d}=\bigcup_{x\in \Gamma_{3d}}\left(\left\{ 2^{3i}\eta \ (\mbox{mod}\ 3r)\ | \ 0\leq i <\frac{n}{3}\right\}\cup\left\{2^{3i+1}\eta \ (\mbox{mod}\ 3r)\ | \ \frac{n}{3}\leq i < \frac{2n-3}{3}\right\}\right).$$
For distinct elements $y_{1}, \ y_{2} \in T_{d}$, if
$$y_{1},\ y_{2}\in\left\{ 2^{3i}\eta \ (\mbox{mod}\ 3r)\ | \ 0\leq i <\frac{n}{3}\right\}$$ or
$$y_{1},\ y_{2}\in\left\{2^{3i+1}\eta \ (\mbox{mod}\ 3r)\ | \ \frac{n}{3}\leq i < \frac{2n-3}{3}\right\},$$ similarly to (i), we have
$3y_{1}\not\equiv3y_{2}\ (\mbox{mod} \ 3r).$ On the other hand, for $2^{3i}\eta,\ 2^{3i^{'}+1}\eta \in T_{d}$ with $0\leq i<n/3$ and $n/3\leq i^{'} <2n/3-1$, if
$$3\cdot2^{3i}\eta\equiv3\cdot2^{3i^{'}+1}\eta\ (\mbox{mod} \ 3r),$$
then $|i-i^{'}|=\frac{n\pm1}{3}$ which contradicts to $3\mid n.$ Therefore,
$$3\cdot2^{3i}\eta\not\equiv3\cdot2^{3i^{'}+1}\eta\ (\mbox{mod} \ 3r).$$
\end{itemize}

It is easy to verify that there do not exist distinct elements $y_{1},\ y_{2}\in T_{d}$ such that $2y_{1}\equiv2y_{2} \ (\mbox{mod}\ 3r)$ or $4y_{1}\equiv4y_{2} \ (\mbox{mod}\ 3r)$. Hence  all of $2T_{d}$, $3T_{d}$ and $4T_{d}$ have the same size as $T_{d}$.

In the following, we will show $iT_{d}\cap jT_{d}=\emptyset$ for distinct $i,j\in[1,4]$. Obviously, $T_{d}\cap 2T_{d}=\emptyset$, $T_{d}\cap 4T_{d}=\emptyset$ and $2T_{d}\cap 4T_{d}=\emptyset$. Note that
$$\mathbb{Z}_{3r}\backslash\{0\}=N_{0}\cup N_{1},$$ where
$$N_{0}=\{x \ \mbox{mod}\ 3r \mid 1\leq x\leq 3r, 3\nmid x\}$$
and
$$N_{1}=\{3x \ \mbox{mod}\ 3r \mid 1\leq x<r\}.$$
Since $3T_{d}\subset N_{1}$ and $j\cdot T_{d}\subset N_{0}$ with $j=1,2,4$, then $T_{d}\cap 3T_{d}=\emptyset$, $2T_{d}\cap 3T_{d}=\emptyset$ and $3T_{d}\cap 4T_{d}=\emptyset$. Hence,
$$M_{4}{'}(3d)\geq|T_{d}|=
\left\{
\begin{array}{ll}
\lfloor\frac{2n}{3}\rfloor\cdot\frac{\varphi(d)}{n}, & \mbox{if} \ 3\nmid n,\\
\frac{(2n-3)\varphi(d)}{3n}, & \mbox{if} \ 3\mid n.
\end{array}
\right.$$

It remains to show the equality holds. We can choose at most one element from the quadruple
$$\{2^{i}\eta,2^{i+1}\eta,2^{i+2}\eta,3\cdot2^{i}\eta\}(\mbox{mod}\ 3r).$$
Suppose that we can choose exactly one element from each quadruple. Firstly, we note that
$$3\cdot2^{i}\eta\equiv3\cdot2^{i+n}\eta\ (\mbox{mod} \ 3r).$$
Then we can not choose $3\cdot2^{i}\eta$. Otherwise, none of $2^{i}\eta,3\cdot2^{i+1}\eta,3\cdot2^{i+2}\eta,2^{i+1+n}\eta,2^{i+2+n}\eta$ can be chosen.
Hence we can only choose $\left\lfloor\frac{2n}{3}\right\rfloor$ elements from the subset
\begin{equation}\label{def U}
U=\{2^{i}\eta(\mbox{mod}\ 3r) \ |\ 0\leq i< n_{1}\}.
\end{equation}
However, if $3\mid n$, then the set
$$\left\{2^{3i}\eta(\mbox{mod}\ 3r) \ |\ 0\leq i< \frac{2n}{3}\right\}$$
contains both $\eta$ and $2^{n}\eta$. Therefore, it is not a valid choice. Hence, in case (ii) we can not find a valid subset of $U$ with size $\frac{2n}{3}$. On the other hand, is is possible to find a valid subset of $U$ in (\ref{def U}) with size $\frac{2n-3}{3}$. Therefore,
$$M_{4}{'}(3d)=
\left\{
\begin{array}{ll}
\lfloor\frac{2n}{3}\rfloor\cdot\frac{\varphi(d)}{n}, & \mbox{if} \ 3\nmid n,\\
\frac{(2n-3)\varphi(d)}{3n}, & \mbox{if} \ 3\mid n,
\end{array}
\right.$$
i.e.,
$$M_{4}{'}(3d)=\left(\left\lceil\frac{2n}{3}\right\rceil-1\right)\cdot\frac{\varphi(d)}{n}.$$

$\mathbf{Case \ 2}$: If $n$ is even, then $n_{1}=n$.

\begin{itemize}
  \item For $d=5$, we have $n=4$ and $M_{4}^{'}(15)=2=2\cdot\left\lfloor\frac{4}{3}\right\rfloor$.
  \item For $d=11$, we have $n=10$ and $M_{4}^{'}(33)=6=2\cdot\left\lfloor\frac{10}{3}\right\rfloor$.
  \item For $d\neq5,11$. It is similar to Case 1. We can choose
$$T_{d}=\bigcup_{x\in \Gamma_{3d}}\left\{ 2^{3i}\eta \ (\mbox{mod}\ 3r)\ | \ 0\leq i <\left\lfloor\frac{n}{3}\right\rfloor\right\}.$$
\end{itemize}\hfill$\blacksquare$\\

 $Example \ 2:$
\begin{itemize}
\item For $q=39$, the construction of maximal set in the proof of Theorem 3 is depicted as follows. Firstly we have $r=13$. Hence, $d=1$ or $d=13$.
If $d=13$, then $n=n_{1}=12$ is even and Case 1(ii) applies. We can choose $\Gamma_{39}=\{1,7\}$ and so
\begin{eqnarray*}
T_{11}&=&\left\{ 2^{3i} \ (\mbox{mod}\ 39) \mid 0\leq i <4 \right\}\cup \left\{ 2^{3i}\cdot7 \ (\mbox{mod}\ 39) \mid 0\leq i <4 \right\}\\
&=& \{1, 8, 25, 5, 7, 17, 19, 35\}
\end{eqnarray*}
is a  maximal size $B_{1}[4](39)$ set.
\item For $q=120$,
$$M_{4}(120)= M_{4}(15)+15.$$
Firstly, the set $\{1,7\}$ is a maximal size $B_{1}[4](15)$ set.
The construction of $B_{1}[4](120)$ set in the proof of Theorem 2 is presented as follows:
\begin{align*}
& S_{0,1}=\{45\},\\
& S_{1,1}=\{35,25\},\\
& S_{0,5}=\{33,39,21,27\},\\
& S_{1,5}=\{31,37,41,43,17,19,23,29\}.
\end{align*}
Therefore
\begin{align*}
& B=S_{0,1}\cup S_{1,1}\cup S_{0,5}\cup S_{1,5}\cup \{8\cdot1,8\cdot7\}=\{8, 17, 19, 21, 23, 25, 27, 29, 31, 33, 35, 37, 39, 41, 43, 45, 56\}
\end{align*}
is a maximal size $B_{1}[4](120)$ set with size 17.
\end{itemize}

\section{Maximal size $B_{1}[4](12r)$ set}

In this section, we give a lower bound of maximal size $B_{1}[4](12r)$
sets from maximal size $B_{1}[4](2r)$ sets with $\gcd(r,6)=1$ and the maximal size $B_{1}[4](2r)$ sets are studied in \cite{XL}. Recall $V_{d}$ in (\ref{def Vd}) and $U_{ij}$ in (\ref{def Uij}).\\

$Theorem \ 4:$ For $\gcd(r,6)=1$,
$$M_{4}(12r)\geq M_{4}(2r)+2r.$$

$Proof:$ For $d\mid r$, let $\alpha=r/d$ and
\begin{align*}
& S_{1}=\left\{i\alpha \ | \ 1\leq i<4d, \ \gcd(i,6d)=1 \right\},\\
& S_{2}=\left\{4r+3i\alpha \ \left| \ 1\leq i<\left\lfloor\frac{2d}{3}\right\rfloor, \ \gcd(i,2d)=1 \right.\right\},\\
& S_{3}=\left\{8r+3i\alpha \ \left| \ \left\lfloor\frac{2d}{3}\right\rfloor+1\leq i<\left\lfloor\frac{4d}{3}\right\rfloor, \ \gcd(i,2d)=1 \right.\right\}.
\end{align*}
Then $S_{i}(i=1,2,3)$ are mutually disjoint by checking the range of values.
Denote by $A_{d}=S_{1}\cup S_{2}\cup S_{3}$
For any $x_{d}\in A_{d}$, \\
$\mathbf{Case \ 1}$: if $x_{d}\in S_{1}$, then
\begin{itemize}
  \item  $2\alpha<2x_{d} \ (\mbox{mod} \ 12r)\ < 8r$ and $4\nmid2x_{d} \ (\mbox{mod} \ 12r)$;
  \item  $3\alpha<3x_{d} \ (\mbox{mod} \ 12r)\ < 12r$ and $9\nmid3x_{d} \ (\mbox{mod} \ 12r)$;
  \item  $4x_{d} \ (\mbox{mod} \ 12r)$ is either in $[4r,12r]$ and not divisible by 8 or in $[1,4r]$.
\end{itemize}
$\mathbf{Case \ 2}$: If $x_{d}=4r+3i\alpha\in S_{2}$ with $1\leq i<\lfloor2d/3\rfloor$, then
\begin{itemize}
  \item  $8r+6\alpha\leq2x_{d} \ (\mbox{mod} \ 12r)=8r+6i\alpha < 12r$ and $8r+6i\alpha\equiv2r \ (\mbox{mod} \ 3)$;
  \item  $9\alpha<3x_{d} \ (\mbox{mod} \ 12r)=9i\alpha < 6r$ and $9\mid3x_{d} \ (\mbox{mod} \ 12r)$;
  \item  $4r+12\alpha\leq4x_{d} \ (\mbox{mod} \ 12r)=4r+12i\alpha < 12r$, $8\mid4x_{d} \ (\mbox{mod} \ 12r)$ and $4r+12i\alpha\equiv r \ (\mbox{mod} \ 3)$.
\end{itemize}
$\mathbf{Case \ 3}$: If $x_{d}=8r+3i\alpha\in S_{3}$  with $\lfloor2d/3\rfloor+1\leq i<\lfloor4d/3\rfloor$, then
\begin{itemize}
  \item  $8r+6\alpha<2x_{d} \ (\mbox{mod} \ 12r)=4r+6i\alpha \ < 12r$ and $4r+6i\alpha\equiv r \ (\mbox{mod} \ 3)$;
  \item  $6r+9\alpha<3x_{d} \ (\mbox{mod} \ 12r)=9i < 12r$ and $9\mid3x_{d} \ (\mbox{mod} \ 12r)$;
  \item  $4r+12\alpha<4x_{d} \ (\mbox{mod} \ 12r)=12i\alpha-4r < 12r$, $8\mid4x_{d} \ (\mbox{mod} \ 12r)$ and $12i\alpha-4r\equiv2r \ (\mbox{mod} \ 3)$.
\end{itemize}
It is easy to see that $|3A_{d}|=|A_{d}|.$
Since $3\nmid r$, then $|2A_{d}|=|A_{d}|$ and $|4A_{d}|=|A_{d}|$.

Moreover, recall $V_{d}$ in (\ref{def Vd}) and $U_{ij}$ in (\ref{def Uij}) (here $a=2$, $b=1$), for any $y=x\frac{r}{d}\in V_{d}\cap U_{01}$, i.e.,
$$x=3k \ \mbox{with} \ 1\leq k<4d \ \mbox{and} \ \gcd(k,2d)=1$$
which implies $y\in3A_{d}$.
Obviously, $3A_{d}\subset V_{d}\cap U_{01}$. Therefore
$$|A_{d}|=|V_{d}\cap U_{01}|=2\varphi(d).$$

Let $A=\bigcup_{d|r}A_{d}.$  Note that
$$A\subset U_{00}, \ 2A\subset U_{10}, \ 3A\subset U_{01} \ \mbox{and} \ 4A\subset U_{20}.$$
Hence $|A|=2r$ and
$$iA\cap (U_{11} \cup U_{21})=\emptyset \ \mbox{for} \ 1\leq i \leq4.$$

We note that if $B_{1}$ is a maximal size $B_{1}[4](2r)$ set, then $6B_{1}$ is a $B_{1}[4](12r)$ set and
$$6B_{1}\subset U_{11}\cup U_{21}.$$
Then $B=A\cup 6B_{1}$ is a $B_{1}[4](12r)$ set of size $M_{4}(2r)+2r$.\hfill$\blacksquare$\\

$Example \ 3:$
\begin{itemize}
\item For $q=60$, we have
$$M_{4}(60)\geq M_{4}(10)+10.$$
It is easy to check that $\{1,9\}$ is a maximal size $B_{1}[4](10)$ set. The construction of $B_{1}[4](60)$ set in the proof of Theorem 4 is presented as follows. Firstly we have $r=5$. Hence, $d=1$ or $d=5$.

If $d=1$, then $\alpha=5$ and $A_{1}=\{5,55\}.$

If $d=5$, then $\alpha=1$ and $A_{5}=\{1,7,11,13,17,19,23,29\}.$

Therefore
$$B=A_{1}\cup A_{5}\cup \{6\cdot1,6\cdot9\}=\{1,5,6,7,11,13,17,19,23,29,54,55\}$$
is a $B_{1}[4](60)$ set with size 12.
\item For $q=84$, we have
$$M_{4}(84)\geq M_{4}(14)+14.$$
By  [\cite{XL}, Theorem 4-(1)], the set $\{1,13\}$ is a maximal size $B_{1}[4](14)$ set.
The construction of a $B_{1}[4](84)$ set in the proof of Theorem 4 is presented as follows. Firstly we have $r=7$. Hence, $d=1$ or $d=7$.

If $d=1$, then $\alpha=7$ and $A_{1}=\{7,77\}.$

If $d=7$, then $\alpha=1$ and $A_{7}=\{1,5,11,13,17,19,23,25,31,37,71,83\}.$

Therefore
$$B=A_{1}\cup A_{7}\cup \{6\cdot1,6\cdot13\}=\{1,5,6,7,11,13,17,19,23,25,31,37,71,77,78,83\}$$
is a $B_{1}[4](84)$ set with size 16.
\end{itemize}

By computer search, $M_{4}(60)=12$ and $M_{4}(84)=16$ which indicates the equality holds in Theorem 4.\\

$\mathbf{Conjecture:}$\ The lower bound of $M_{4}(12r)$ deduced from Theorem 4 is tight for $\gcd(r,6)=1.$

 \section{Maximal size $B_{1}[4](2\cdot3^{b}r)$ set}

In this section, we give a lower bound of maximal size $B_{1}[4](2\cdot3^{b}r)$
set from a maximal size $B_{1}[4](2\cdot3^{b-3}r)$ set with $\gcd(r,6)=1$. Recall $U_{ij}$ in (\ref{def Uij}).\\

$Theorem \ 5:$ If $b\geq3$ and $\gcd(r,6)=1$, then
$$M_{4}(2\cdot3^{b}r)\geq M_{4}(2\cdot3^{b-3}r)+8\cdot3^{b-3}r-\sum\limits_{\substack{d\mid r\\
3^{b-2}\nmid \ \text{ord}_{d}(2)}}\frac{2\cdot3^{b-3}\varphi(d)}{\mbox{lcm}(2\cdot3^{b-3},\text{ord}_{d}(2))}.$$

$Proof:$ Let $\alpha=r/d$ and
\begin{eqnarray*}
A_{d}\!&=&\!\left\{i\alpha \ | \ 1\leq i\leq1+3^{b-1}d, \ \gcd(i,2\cdot3^{b-1}d)=1 \right\}\\
&&\cup\left\{i\alpha+4\cdot3^{b-1}r \ | \ 1+3^{b-1}d< i<2\cdot3^{b-1}d, \ \gcd(i,2\cdot3^{b-1}d)=1 \right\}.
\end{eqnarray*}
Note that $|A_{d}|=\varphi(2\cdot3^{b-1}d)=2\cdot3^{b-2}\varphi(d)$. Denote by $A=\bigcup_{d|r}A_{d}$. Then
$$|A|=\sum_{d|r}|A_{d}|=\sum_{d|r}2\cdot3^{b-2}\varphi(d)=2\cdot3^{b-2}r.$$

It suffices to shows $A$ satisfies $(1.1)$. For any $x\in A$, we have
$$3\nmid x \ \mbox{and} \ A\cap 3A=\emptyset.$$
Checking binary parity we can get $A\cap 2A=\emptyset, \ A\cap 4A=\emptyset, \ 2A\cap 3A=\emptyset$ and $3A\cap 4A=\emptyset$.
Clearly, if $d\neq d^{'}$ with $d\mid r$ and $\ d^{'}\mid r$, then
$$2A_{d}\cap 4A_{d^{'}}=\emptyset.$$
Moreover, the following statements hold for any $x_{d}\in A_{d}$,\\
1) $2\alpha<2x_{d}\leq 2\alpha+3^{b-1}r$ and $4\nmid2x_{d}$;\\
2) $2\alpha+4\cdot3^{b-1}r<2x_{d} \ (\mbox{mod} \ 2\cdot3^{b}r)\leq 2\cdot3^{b}r$ and $4\mid2x_{d} \ (\mbox{mod} \ 2\cdot3^{b}r)$;\\
3) $4\alpha<4x_{d} \ (\mbox{mod} \ 2\cdot3^{b}r)\leq 4\alpha+4\cdot3^{b-1}r$ and $4\mid4x_{d} \ (\mbox{mod} \ 2\cdot3^{b}r)$;\\
4) $4\alpha+4\cdot3^{b-1}r<4x_{d} \ (\mbox{mod} \ 2\cdot3^{b}r)\leq 2\cdot3^{b}r$ and $4\nmid4x_{d} \ (\mbox{mod} \ 2\cdot3^{b}r)$.\\
It follows that $2A_{d}\cap4A_{d}=\emptyset$ and so $2A\cap4A=\emptyset$.

Let $m_{i}=\mbox{ord}_{3^{i}d}(2)$ and $\Gamma_{d}$ be a set of coset representatives of the group generated by 2 in $\mathbb{Z}_{3^{b-1}d}$.\\
By $[7, \ Th.10]$, the set
$$S=\bigcup\limits_{d\mid r}6T_{d}$$
is a $B_{1}[4](2\cdot3^{b}r)$ set where $T_{d}$ is depicted as follows:
\begin{itemize}
  \item if $m_{b-1}=3m_{b-2} \ \mbox{and} \ 3\nmid m_{b-2}$, then
  $$T_{d}=\bigcup\limits_{x\in\Gamma_{d}}\left\{2^{3i}xr/d \ (\mbox{mod} \ 3^{b-1}r) \ \big| \ 0\leq i< m_{b-2}\right\};$$
  \item if $m_{b-1}=3m_{b-2} \ \mbox{and} \ 3\mid m_{b-2}$, then
  \begin{align*}
  T_{d}= & \bigcup\limits_{x\in\Gamma_{d}}\Big(\left\{2^{3i}xr/d \ (\mbox{mod} \ 3^{b-1}r) \ \big| \ 0\leq i< m_{b-2}/3\right\}\\
  & \cup\left\{2^{3i+1}xr/d \ (\mbox{mod} \ 3^{b-1}r) \ \big| \ m_{b-2}/3\leq i< 2m_{b-2}/3\right\}\\
  & \cup\left\{2^{3i+2}xr/d \ (\mbox{mod} \ 3^{b-1}r) \ \big| \ 2m_{b-2}/3\leq i< m_{b-2}-1\right\}\Big);
  \end{align*}
  \item if $m_{b-1}=m_{b-2}$, then
  $$T_{d}=\bigcup\limits_{x\in\Gamma_{d}}\bigcup\limits_{\delta=0}^{2}\left\{2^{3i+\delta}(xr/d+3^{b-2}r\delta) \ (\mbox{mod} \ 3^{b-1}r) \ \big| \ 0\leq i< m_{b-1}/3\right\}.$$
\end{itemize}
Then $S\subset U_{11}$ with size $2\cdot3^{b-3}r-\sum\limits_{\substack{d\mid r\\
3^{b-2}\nmid \ \text{ord}_{d}(2)}}\frac{2\cdot3^{b-3}\varphi(d)}{\mbox{lcm}(2\cdot3^{b-3},\text{ord}_{d}(2))}$.

Let $S_{b}$ be a $B_{1}[4](2\cdot3^{b-3}r)$ set. Define
$$B=A\cup S\cup S^{'} \ \mbox{with} \ S^{'}=\{27c \mid c\in S_{b}\}.$$
It is easy to verify that
$B$ is a $B_{1}[4](2\cdot3^{b}r)$ set and
$$|B|=M_{4}(2\cdot3^{b-3}r)+8\cdot3^{b-3}r-\sum\limits_{\substack{d\mid r\\
3^{b-2}\nmid \ \text{ord}_{d}(2)}}\frac{2\cdot3^{b-3}\varphi(d)}{\mbox{lcm}(2\cdot3^{b-3},\text{ord}_{d}(2))}.$$
\hfill$\blacksquare$\\

$Example \ 4:$
For $q=54$, we have
$$M_{4}(54)\geq M_{4}(2)+8=8.$$
The construction of a $B_{1}[4](54)$ set in the proof of Theorem 5 is presented as follows:
$$ A=\{1,5,7,47,49,53\}, \ S=\{6,48\} \ \mbox{and} \ S^{'}=\emptyset.$$
Therefore
$$B=A\cup S=\{1,5,6,7,47,48,49,53\}$$
is a $B_{1}[4](54)$ set with size 8 but it is not maximal. By computer search,
$$\{1, 5, 7, 8, 9, 40, 46, 49, 51\}$$
is a maximal size $B_{1}[4](54)$ set with size 9.

\section{Summary}

In this paper, we are mainly consider the constructions of a maximal size $B_{1}[4](2^{a}3^{b}r)$ set with $\gcd(r,6)=1.$ It can be applied to error correction for single asymmetric error of limited magnitude since all the syndromes are distinct. Firstly, the construction in Theorem 1 is elementary. Note that $S_{2}=-S_{1}$ in the proof of Theorem 1 which implies that it can apply to the $B_{1}[\pm4](2^{a}3^{b}r)$ set. Moreover, $B_{1}[\pm4](2^{a}3^{b}r)$ set can be used to study equi-difference conflict-avoiding code with weight 5.
For $a\equiv0 \ (\mbox{mod} \ 3)$ and $b\equiv1 \ (\mbox{mod} \ 2)$, combining Theorems 1-3 and [\cite{KLNY}, Th.10],  we have completely determined maximal size $B_{1}[4](2^{a}3^{b}r)$ sets with $\gcd(r,6)=1$. Furthermore, we discuss maximal size $B_{1}[4](12r)$ sets and give a lower bound of $M_{4}(12r)$ from maximal size $B_{1}[4](2r)$ sets and the maximal size $B_{1}[4](2r)$ sets are studied in \cite{XL}. Finally, we give a lower bound of maximal size $B_{1}[4](2\cdot3^{b}r)$
set from a maximal size $B_{1}[4](2\cdot3^{b-3}r)$ set. It may not always be tight since we do not choose elements from $U_{02}$.


\end{document}